\documentstyle[12pt,aasms4]{article}

\begin{document}

\title{Simulated VLBI Images From Relativistic Hydrodynamic Jet Models}
\author{Amy J. Mioduszewski, Philip A. Hughes}
\affil{Department of Astronomy, University of Michigan, Ann Arbor, MI 
48109-1090}
\authoremail{amy@astro.lsa.umich.edu}
\and
\author{G. Comer Duncan}
\affil{Department of Physics and Astronomy, Bowling Green State University, 
Bowling Green, OH 43403}

\begin{abstract}
A series of simulated maps showing the appearance in total intensity of flows
computed using a recently developed relativistic hydrodynamic code (Duncan \&
Hughes 1994: ApJ, 436, L119) are presented.  The radiation transfer
calculations were performed by assuming the flow is permeated by a magnetic
field and fast particle distribution in energy equipartition, with
energy density proportional to the hydrodynamic energy density (i.e.,
pressure).  We find that relativistic flows subject to strong perturbations
exhibit a density structure consisting of a series of nested bow shocks, and
that this structure is evident in the intensity maps for large viewing
angles. However, for viewing angles $<30^{\circ}$, differential Doppler
boosting leads to a series of axial knots of emission, similar to the pattern
exhibited by many VLBI sources.  The appearance of VLBI knots is determined
primarily by the Doppler boosting of parts of a more extended flow.  To study
the evolution of a perturbed jet, a time series of maps was produced and an
integrated flux light curve created.  The light curve shows features
characteristic of a radio loud AGN: small amplitude variations and a large
outburst. We find that in the absence of perturbations, jets with a modest
Lorentz factor ($\sim 5$) exhibit complex intensity maps, while faster jets
(Lorentz factor $\sim 10$) are largely featureless.  We also study the
appearance of kiloparsec jet-counterjet pairs by producing simulated maps at
relatively large viewing angles; we conclude that observed hot spot emission
is more likely to be associated with the Mach disk than with the outer, bow
shock.
\end{abstract}

\section{Introduction}

Most radio loud Active Galactic Nuclei (AGN), when mapped using Very Long
Baseline Interferometry (VLBI), show a stationary core and knots of emission
which sometimes move superluminally.  These features are believed to be the
most prominent parts of a jet of relativistic plasma (Blandford \& K\"{o}nigl
1979; Lind \& Blandford 1985), and the superluminal motion and general
absence of a counterjet leave little doubt that relativistic effects (Doppler
shift, Doppler boosting, aberration, time delays, etc.) play a crucial role
in determining the appearance of these flows.  The recent development of
relativistic hydrodynamic codes has greatly enhanced our ability to explore
the dynamics of such extragalactic jets. However, a comparison of the
simulated flow with single-dish and VLBI data requires the computation of a
radiated flux -- ideally, the Stokes parameters $I$, $Q$ and $U$ -- from the
flow. Since the relativistic effects influencing the appearance of AGN are
strongly dependent on the viewing angle, it is important to compute simulated
images for a jet aligned at various angles to the line of sight.  In some
earlier studies this has been done using non-relativistic hydrodynamics to
simulate the conditions inside the jet and relativistic formulae to compute
the radiation flux. For example, Wilson and Scheuer (1983) study the
appearance of kiloparsec scale structures assuming that no relativistic
particles (i.e. synchrotron emission) are present except in the shock front.
With the advent of relativistic hydrodynamical codes a consistent calculation
of the radiation field is possible which allows one to properly include
relativistic effects in both dynamics and radiation transfer.

As the radiated flux is determined by the magnetic field and fast particle
distributions, which are not computed in the hydrodynamic simulations, some
assumptions must be made about how those quantities are related to the
hydrodynamic variables.  Our assumptions about these quantities (namely, that
the magnetic field and particle distributions are proportional to the
hydrodynamical pressure) are very similar to those adopted in
non-relativistic or steady state relativistic  simulations (e.g. Rayburn
1977; Williams \& Gull 1984; Wilson 1987).  Jun, Clark \& Norman (1994) have
studied cosmic-ray mediated magnetohydrodynamical shocks using a two-fluid
approach; such a consistent computation of the magnetic field and radiating
particle distributions is highly desirable, but is currently beyond the
scope of studies that aim to elucidate the overall flow dynamics of
relativistic jets.   In this paper we present the first results from a study
that adopts a simple mapping between hydrodynamic and high energy species,
and that highlights the very different morphologies exhibited by the flow
material and by the associated radiation flux.  In \S 2, the relativistic
hydrodynamical code used to produce the data is summarized.  \S 3 describes
the radiation transfer calculations.  We discuss the simulated maps and the
integrated flux light curve in \S 4.  Conclusions and future work are
presented  in \S 5.

\section{Relativistic CFD}

We use a method for the numerical solution of the Euler equations which has
been found to be both robust and efficient, and which permits treatment of
relativistic flows. The evolved variables are mass, momentum and total energy
density, in the {\it laboratory} frame. With the adoption of these
quantities, the relativistic Euler equations have a form identical to that of
the nonrelativistic equations, thus allowing a direct application of
techniques devised for the latter. Our approach employs a solver of the
Godunov-type, with approximate solution of the local Riemann problems.  In
this method, the RHLLE technique, a relativistic generalization of a method
originally developed for non-relativistic fluids (Harten {\it et al.} 1983;
Einfeldt 1988), the full solution to the Riemann problem is approximated by
two discontinuities separating a constant state, whose value must satisfy the
Euler equations in conservation form.  However, velocity and pressure appear
explicitly in the relativistic Euler equations, in addition to the evolved
variables, and pressure and rest density are needed for the computation of
the wave speeds that form the basis of the numerical technique. We obtain
these values by performing a Lorentz transformation at every time and cell
boundary (or center) where the rest frame values are required. The Lorentz
transformation involves a numerical root finder to solve a quartic equation
for the velocity.  This provides robustness, because it is straight forward
to ensure that the computed velocity is always less than the speed of light.
The relativistic Euler equations and Lorentz transformation are described in
Appendix A.

We achieve second-order accuracy in time by computing fluxes at the half-time
step. The Godunov method requires a `reconstruction' step, in which
cell-centered values of variables in juxtaposed cells are used to estimate
the cell boundary values of these quantities. It is this linear interpolation
that provides second-order spatial accuracy. However, it is possible
for the rest frame quantities corresponding to the interpolated laboratory
frame values to be aphysical, corresponding to a velocity in excess of the
speed of light, and/or negative pressure. Such behavior is easily trapped,
and our scheme (rarely) falls back locally to first-order if needed.

The solver is implemented in a 2-D axisymmetric form within the framework of
an Adaptive Mesh Refinement algorithm (Quirk 1991), allowing us to perform
high-resolution, 2-D simulations with modest computing resources.  Adaptive
Mesh Refinement is used to ensure that the grid density is locally adequate
for an accurate rendition of sharp features, such as shocks, while admitting
computations on workstation-class machines of modest speed and memory.  In
this approach the solution is stored in a hierarchy of `patches', each of
which is a logically rectangular grid, with a number of patches at each level
of the hierarchy. In regions of little activity, a coarse grid is sufficient,
and the solution is known on a set of abutting domains with cell size equal
to that adopted for the unrefined grid. In regions where significant
structure lies, the solution must be taken from patches of higher cell
density, embedded within the coarsest mesh. One must either interpolate the
values of the state variables to a uniform mesh with scale equal to that of
the finest refined mesh (with consequent increase in needed storage) or, when
performing the radiation transfer calculations discussed herein, compute the
locations of intersection between a line-of-sight and the boundaries that
describe the hierarchy of patches. The latter is particularly difficult to
implement when time delay effects are to be considered, because the data
populating the cells used for radiation transfer will be epoch-dependent at
each point along a ray, but the patch structure changes with epoch. We are
currently building a radiation transfer code that can accommodate time delays
without losing the benefits of the Adaptive Mesh structure; however, for the
radiation transfer calculations described here, we interpolate onto a single
fine mesh.  Combinations of various Lorentz factors ($1<\gamma<10$), Mach
numbers ($6<\fam2 M<$ $15$) and adiabatic indices ($\Gamma=4/3$ or
$\Gamma=5/3$) were studied by Duncan \& Hughes (1994), where some further
details of both the solver and Adaptive Mesh Refinement are presented,
together with the first results from this code.

\section{Radiation Transfer Calculations}

For the radiation transfer calculation, the potentially complex mesh
structure associated with AMR is circumvented, by first interpolating the
hydrodynamic data onto a single fine rectangular mesh, that represents a cut
through the axis of the axially symmetric flow. The scale of this mesh is
chosen to equal that of the most refined patches employed in the hydrodynamic
simulation.  A Lorentz transformation of the values determined by the
hydrodynamic simulation then provides the pressure and the axial and radial
components of velocity in each cell of this mesh.  The rectangular data set
may be rotated through $360^{\circ}$ to populate a cylindrical volume with 3-D
data.  To facilitate the radiation transfer calculations, a 3-D rectangular
coordinate system is established, so that  the `observer's' lines of sight
are in planes parallel to one side of the mesh. This mesh is populated with
data by reference to the original, fine 2-D data set, having computed the
axial and radial coordinates corresponding to the cell location in three
dimensions. The radial component of velocity is decomposed into two Cartesian
components, thus providing each cell with a value for pressure and three
components of velocity.  For a given viewing angle, lines of sight are
projected through the mesh.  For each intersected mesh cell the hydrodynamic
pressure and velocities are extracted from the 3-D data set.  Then for each
line of sight, starting at the far side of the mesh and stepping along the
line of sight, these values are used in radiation transfer calculations to
determine the flux at the surface of the mesh.   The mesh scale can be
coarsened for a `quick look' at the data set, or used at a scale similar to
that of the data.  The viewing angle, optical depth, spectral index and
frequency are free parameters in a given radiation transfer computation.  
With no coarsening, the mesh is 2000 by 500 by 500 cells for the cases explored
here.

A primary goal of this study is to examine the distribution of synchrotron
flux; thus values for
the synchrotron emissivity and opacity must be computed.  The synchrotron
emissivity ($j_\nu$) and opacity ($\kappa_\nu$) depend upon the high energy
particle distribution ($n_0$), the magnetic field ($B$), the frequency
($\nu$), and the spectral index ($\alpha$).  In the reference frame of the
plasma (Pacholczyk 1970):

\begin{equation}
j_\nu \propto n_0 B^{\alpha+1}\nu^{-\alpha},
\end{equation}
\begin{equation}
\kappa_\nu \propto n_0 B^{(\alpha+\frac{3}{2})}\nu^{-(\alpha+\frac{5}{2})}. 
\end{equation}

Assuming minimum energy, which approximates energy equipartition, it follows
that the radiating particle number density, $n_0$, and magnetic field energy
density, $u_B$, are directly proportional to the hydrodynamical pressure
($p$), i.e., the internal energy density. If
\begin{equation}
u_B=u_e,
\end{equation}
\noindent and
\begin{equation}
p=(\Gamma-1)u_e,
\end{equation}
\noindent where $\Gamma$ is the adiabatic index, then $u_B\propto p$ and
$u_e\propto p$.  Therefore, since the magnetic field energy density is
\begin{equation}
u_B=\frac{B^2}{2\mu_0}
\end{equation}
\noindent then 
\begin{equation}
B\propto p^{\frac{1}{2}}.
\end{equation}
\noindent The high energy particle energy density is given by
\begin{equation}
u_e=\int_{E_L}^{E_H} n_0 E^{-\delta} E dE,
\end{equation}
\noindent where $E_H$ and $E_L$ are the high and low particle energy cutoffs
and $\delta$ is the slope of the particle energy spectrum ($=2\alpha+1$).
If $E_H \gg E_L$ and $\delta > 2$, then
\begin{equation}
u_e \simeq \frac{n_0 E_{L}^{-\delta+2}}{\delta-2}
\end{equation}
\noindent so that
\begin{equation}
n_0 \propto p
\end{equation}
if $E_L$ is a constant, or slowly varying function of position and time.

\noindent So, substituting equations (6) and (9)  into equations (1) and (2)
we find:

\begin{equation}
j_\nu \propto p^{\frac{\alpha+3}{2}}\nu^{-\alpha},
\end{equation}
\begin{equation}
\kappa_\nu \propto p^{\frac{2\alpha+7}{4}}\nu^{-(\alpha+\frac{5}{2})}.
\end{equation}

Equations (10) and (11) describe the dependency of the radiation transfer
coefficients on pressure, but some normalization must be adopted. That of the
emissivity is arbitrary, as the underlying hydrodynamic calculations were
independent of length scale and an arbitrary choice of length scale would
lead to an arbitrary intensity for an optically thin flow. The normalization
of opacity is chosen to provide the adopted optical depth (which is a free
input parameter) for a line of sight with a typical path length through the
flow at the given angle of view.  An average value of the minimum and
maximum pressures ($<p>$) over the whole computational domain is used in
computing the normalization.

Adopting $\nu=1$ as a fiducial frequency, and with $L$ the total path length
through the flow as just described, for a desired optical depth $\tau$
\begin{equation}
\kappa_\nu = \frac{\tau}{L} \left(\frac{p}{<p>}\right)^{\frac{2\alpha+7}{4}}
\nu^{-(\alpha+\frac{5}{2})}.
\end{equation}
The actual optical depths for different lines of sight deviate from $\tau$,
but this approach provides a method of tuning the optical depth to explore
the appearance of optically thin and thick flows, through a single parameter
that may be set prior to performing the radiation transfer calculations.

Synchrotron radiation transfer calculations for the total intensity, $I$, are
performed following Rybicki \& Lightman (1979), allowing for Doppler boost
and frequency shift.  Generally, the laboratory frame spectral intensity,
$I$, is equal to ${\fam2 D}^3$ times the rest frame spectral intensity, where
${\fam2 D}=(\gamma(1-\beta cos\theta))^{-1}$ is the Doppler factor for a flow
speed $\beta c$ and angle of view $\theta$.  However, as we discuss below,
the pattern of structures evident in the flows under study changes slowly, and
we can approximate the flow as a fixed distribution of relativistic
velocities within a stationary `window' in the observer's frame. Thus
(Cawthorne 1991)

\begin{equation}
I=I_0e^{-l \kappa_\nu}+\frac{j_\nu}{\kappa_\nu}{\fam2 D}^2(1-e^{-l 
\kappa_\nu}) .
\end{equation}

\noindent where $\beta$ is the velocity normalized to the speed of light,
($v/c$),  $\gamma=(1-\beta^2)^{\slantfrac{-1}{2}}$ and $l$ is the line of
sight thickness of an individual cell.  We have used $\alpha=0.75$ throughout
the computations reported therein.

The magnetic field is assumed to be tangled with length scale much less than
that of a computational cell.  Therefore, since for the simulations reported
here, there is no preferred field direction within a cell, aberration does
not change the average effective field orientation and may be ignored.  Time
delay effects have been ignored because although the maximum instantaneous
flow speeds are relativistic, the jet {\it structures} move at barely
relativistic speeds:  we observe $\beta_{shock}\simeq 0.59$ for the slowest
relativistic case and $\beta_{shock}=0.89$ for the fastest relativistic case
studied here. This is a consequence of the large amplitude variations in
inflow Lorentz factor for the perturbed case: the shocks driven by these
variations are strong, and so move rapidly forward in the frame of the
upstream fluid; weak shocks would move at close to the fluid speed, and thus
move rapidly in the observer's frame.  In fact, the exclusion of such delay
effects does not impact our conclusions which depend on the significant
differential Doppler boosting between flow close to, and flow far from the
axis, not on placement of structures along the flow axis.

\section{Maps and Analysis}

Figure~1 contains schlieren-type images which shows the {\it gradient} of the
laboratory frame density from the hydrodynamical simulations used to produce
the images shown in Figures 2-6. The first four cases are from Duncan \&
Hughes (1994). The first three have Lorentz factors $\sim 1$, $5$ and $10$
respectively, and adiabatic index $5/3$; the fourth has Lorentz factor $10$,
and adiabatic index $4/3$.  The fifth case has the same parameters as the
fourth, but the inflow Lorentz factor was sinusoidally modulated between
$\sim 1$ and $10$ to induce perturbations.  Figures 2-6 present the results
of the radiation transfer calculations for these five hydrodynamic
simulations at four viewing angles ($\theta$):  $10^{\circ}$, $30^{\circ}$,
$60 ^{\circ}$ and $90^{\circ}$. The four panels are logarithmically-scaled
contour maps produced by the radiation transfer program at the four viewing
angles.  The peak flux differs from map to map, and is proportional to the
Doppler boost except in the nonrelativistic case (Figure~2).  Note that the
maps are dominated by the head/bow region and while the axial
incident-reflection shock pattern remains visible in the nonrelativistic
case, in general there is little internal jet structure evident.  The dynamic
range of these maps is similar to a low (20:1) dynamic range VLBI map.

The hydrodynamic simulation of the perturbed jet exhibits a series of nested
bow shocks, which are also evident in the simulated maps (Figure~6) for
angles of $90^{\circ}$ and $60^{\circ}$; however when the viewing angle is
decreased to less than $60^{\circ}$ the pattern of emission takes the form of
{\bf axial knots}.  To explore this effect, Figure~7 presents a calculation
of the emissivity, and the Doppler boosting ($j\propto p^\frac{\alpha+3}{2}$,
${\fam2 B}\propto {\fam2 D}^{2+\alpha}$) at four angles of view for a slice
through the center of the perturbed jet.  Figure~7a.  shows that rest frame
emissivity is enhanced primarily at the bow shocks.  Figures~7b.-7c.
demonstrate that at small viewing angles the Doppler boosting accentuates the
core of the flow, while at larger viewing angles Doppler boosting has little
effect on the appearance of the jet.  Comparing the distinctive signature of
emissivity and Doppler boosting shown in Figure~7 to the morphology of the
maps shown in Figure~6 strongly suggests that at small viewing angles the
image morphology is determined primarily by the Doppler boosting of the high
velocity jet, whereas at larger angles the intrinsic emissivity is more
important.

To examine where the transition in viewing angle from `bow shock dominated'
to `jet flow dominated' flow occurs, a simple measure of the local flux
contribution (${\fam2 B}  j$) was calculated for a patch near the jet axis
and a patch off axis containing  bow shock structure, as a function of
viewing angle. The results of this calculation, shown in Figure~8,
demonstrate that the angle at which transition from `bow shock dominated' to
`jet flow dominated' occurs is $\sim 20^\circ$. This issue may also be
addressed analytically (see Appendix B), by relating emissivity to pressure,
and pressure to the velocity jump at the bow shock, and determining at what
angle of view Doppler boosting of the jet flow produces an intensity that
exceeds that associated with the bow a few jet radii off axis. This approach
leads to a similar conclusion, namely that the jet should dominate for
viewing angles less than $\sim 30^\circ$.  These results are somewhat at odds
with that from visual inspection of the maps, which indicates that the
transition is somewhere between $30^\circ$ and $60^\circ$.  The explanation
of this lies in the fact that neither calculation take line of sight effects
into account.  The regions of high Doppler boosting are `thick' ($\sim 120$
pixels wide) as opposed to the regions of high emissivity which are long and
`thin' (only $\sim 15$ pixels wide).  Therefore, at large viewing angles a
line of sight will travel through more cells of high emissivity than at small
angles.  The exact opposite would be true for the regions of high Doppler
boosting. Taking this into account, the local flux contribution as a
function of viewing angle was calculated by summing along a line of sight
through 2 patches, one containing an off-axis bow shock and the other 
containing a region of high Doppler boosting near the axis, both
approximately 120 pixels wide.  The line of sight, for the patch containing
the bow shock, was selected so that it would intersect the bow shock for
all angles of view.  These calculations indicate a transition angle of
$\sim 50^\circ$ (see Figure 9) which is much more consistent with what
is shown in the maps.

For the perturbed case, maps were generated using output of the
hydrodynamical data every 150 computational cycles, to create 26 time slices
of the perturbed jet. The intensities were summed for each map, and a light
curve such as would result from single dish monitoring was created at three
different frequencies.  These light curves are shown in Figure~10 for a
viewing angle of $30^{\circ}$; the `central' frequency ($\nu$ in Figure~10)
is the frequency used in previously discussed simulations. The simulated
light curves are indeed suggestive of some of the large amplitude outbursts
displaying substructure, constant flux level, seen in the University of
Michigan Radio Astronomy Observatory (UMRAO) database (Aller {\it et al.}
1985). An example of this is shown in Figure~11, where beginning in 1988,
UMRAO observed an outburst in the BL Lac object 0735+178 (Aller, H. D. \&
Aller, M. F.  private communication).  As in the simulated light curve, there
are low amplitude fluctuations as well as a large amplitude outburst.  The
variations were examined in detail by calculating the emissivity and Doppler
boosting along the jet axis for all time slices.  This shows that the lower
amplitude total flux variations are indeed a result of the onset of shocks.
These simulations cannot of course be expected to reproduce many of the
features seen in single dish monitoring data, which are generally accepted to
arise from the passage of shocks into an optically thin portion of a
diverging flow, with subsequent adiabatic energy loss, because the
hydrodynamic models employed here have neither a diverging inflow nor an
ambient pressure gradient, and so undergo no significant lateral expansion.

A further similarity to the monitoring data is the damping of the variations
at the lowest frequency.  The nature of the variations seen in the lowest
flux curve ($\nu/3$ in Figure~10) is explained by the fact this is near the
spectral turnover between the optically thin and optically thick parts of the
spectrum, and opacity effects are masking the contributions from far portions
of the flow:  we see only the longish time scale fluctuations of structures
near to the $\tau=1$ surface, rather than the sum of the weakly correlated
variations from the whole body of the emitting volume.  In contrast, the
structure in the higher frequency light curves is caused both by the creation
of new components at the inflow and by the merging of components.  A striking
feature of the light curves is that there is very little evidence for
periodicity, which is surprising given that the perturbations were driven
using a sinusoidal modulation of the inflow Lorentz factor. To ascertain
whether a Fourier analysis could pick out periodicity where the eye could
not, we constructed a Scargle periodogram (Scargle 1982), shown in
Figure~12.  (Scargle provides a `false alarm probability', which aids in
judging the significance of peaks in the power distribution.) No periodicity
is evident, there being only the broad distribution of power associated with
the large amplitude rise seen in the light curves. Evidently such a feature,
occurring within a time series with limited sampling of only a few cycles of
the modulation, masks the signature of the latter. This may be a warning that
to see clear evidence of periodicity, it is necessary to have a well-sampled
data set spanning many cycles of activity sustaining the same frequency of
variation. Indeed, the character of radio waveband variability can change
significantly over a time scale of years and there is little evidence from
such data sets for periodicity (Aller, Aller \& Hughes 1996).

Figure~13 shows simulated maps corresponding to the 11-14th time slices,
which cover the time interval of the large outburst seen in the light curve.
Notice that in Figure~13a and b, the 1st and 2nd components merge, and a 4th
knot is formed.  Also note that the components in these maps move away from
the core -- behavior very similar to that seen in many multi-epoch VLBI maps
(e.g., Zensus \& Pearson 1990).  Figure~15 shows the motion of each component
over time.  The position of each component was determined by recording the
position of the peak flux.  The components move at about the same speed, the
average component velocity being $0.14$ jet radii/time step with a standard
deviation of $0.033$ jet radii/time step.  Also notice the apparent
acceleration in component 3 between time steps 20 and 21. Component 3 is
double peaked and between time steps 20 and 21 the maximum flux in the
component moves from the rear-most peak to the forward-most peak, causing
this apparent acceleration.  This would suggest
that component accelerations in
observed jets might be associated with a continuous and simple change in the
distribution of emitting material, rather than acceleration of plasma, or a
change in shock propagation speed. It must be noted that we have not
seen merging produce motion against the flow.

A plot of the fluxes of the individual components as they move along the jet
is shown in Figure~15.  Most components show similar flux histories as they
evolve.  However, the first and `combined' components show dramatic deviation
from this behavior.  The flux increase in the `combined' component dominates
the outburst shown in the light curve.  An examination of the component
fluxes verses time shows that the outburst arises from the fact that the
merged component has $\sim 2.5$ times the flux as the summed fluxes of
components 1 and 2 in the previous time slice.  Inspection of the
hydrodynamic variables shows that this outburst results from a dramatic
increase in the hydrodynamical pressure at the merger site -- the dramatic
rise in flux is a direct consequence of hydrodynamical effects, not of,
for example, a change in Doppler boosting associated with a change in flow
speed.

\section{Kiloparsec Scale Structures}

Although we have focused on the `VLBI-like' structure it is important to note
that the hydrodynamic simulations are scale-free and therefore can be applied
to the study of kiloparsec scale jets.  There has long been a debate about
the origin of the emission from the `hot spot and nearby lobe' structure seen
in many FR II radio galaxies (Bridle \& Perley 1984).  The emission from the
hot spot and its environment could originate from an internal shock (the Mach
disk), from the more extended bow shock, or from both structures.  Recall
that we assume emissivity is dependent only on pressure, i.e., particle
acceleration is not included in our modeling. The distribution of pressure
maxima and minima in the vicinity of the Mach disk-bow shock region -- and
thus the distribution of rest frame emissivity -- will not be the same as the
distribution of particles accelerated by, for example, the first-order Fermi
process at either the Mach disk or the bow shock. If the emission is
determined by the latter, then maps produced using the techniques discussed
above, although providing a valid indication of the likely general
distribution of intensity, will not be right in detail. With that caveat in
mind, we now ask how the intensity distribution correlates with flow
structures in the head of the source.

Figure~16 shows maps of two-sided jets with intermediate viewing angles
($45^\circ$, $60^\circ$, and $75^\circ$) using the $\gamma=10$ simulation
(see Figures 1d. and 5).  The jet that is approaching is on the right while
the corresponding receding jet is on the left.  Notice that the maps in
Figure~16 show a hot spot and nearby lobe structure, the latter becoming more
prominent with an increase of viewing angle.  Also, as the viewing angle
increases the opposing jet becomes stronger and a lobe structure develops
there as well.  The brightness ratios between the pairs of jets are 19.4, 6.9
and 2.5 for the $45^{\circ}$, $60^{\circ}$, $75^{\circ}$ jet pairs
respectively.  Bridle and Perley (1984) defined `one-sided' jets as those
with a brightness ratio of greater than 4:1, using this definition
$45^{\circ}$ and $60^{\circ}$ pairs of jets can legitimately be defined as
one-sided, which is in accordance with the view that classical doubles are
twin-jet sources seen at angles of view substantially larger than $45^{\circ}$.

Figure~17 shows a grey scale schlieren-type image of the pressure gradient at
the head of the jet, with a plot of the flux in a slice of the jet as
contours superimposed upon it.  The Mach disk is the structure perpendicular
to the flow axis, and is marked on the grey scale by the arrow.  The peak of
emission -- which we would call the hot spot -- is associated with the Mach
disk and the material just down stream from the Mach disk.  The more diffuse
lobe structure is associated with shocked ambient medium, just down stream
from the bow shock.

We conclude that localized regions of emission near to the periphery of FR II
sources are likely to be associated with thermalization of the jet flow at a
Mach disk, and may thus be used to infer the pressure down stream of that
structure, which may in turn be used to build simple analytic models for the
flow dynamics (e.g., Williams 1991). The measured spectral properties of this
emission are providing information on processes in shocked {\it jet} material,
the composition of which is still debated, at point where little entrainment
has occurred as the jet is cocooned for most of its length.

\section{Conclusions and Future Work}

The simulated images of quiescent flows shows that for Lorentz factors in
excess of $\sim 5$, little structure is evident within the jet. However,
images of a perturbed relativistic jet seen close to the line of sight show a
sequence of resolved axial knots similar to those seen on VLBI maps.  It is
striking that although the morphology of the hydrodynamic quantities is very
different from the morphology of observed jets, for small viewing angles,
relativistic effects dominate to produce images that closely resemble the
observations.

In the simulations using periodic perturbations the associated light curves,
whose appearance is determined by the onset and merging of individual
`events', do not reflect the periodic nature of the perturbations.  In fact,
it was very difficult to relate the features of a light curve to the flux
maps without detailed examination of the hydrodynamic simulations used to
make them: there is a complex relation between the maps and the underlying
flow morphology. In particular, a flux outburst associated with one of the
components was the result of the interaction of two flow structures, and a
detailed study of component motion revealed that apparent (slight)
accelerations can be the result of a change in the spatial distribution of
hydrodynamical quantities, rather that a simple acceleration of plasma or the
onset of shocks.

The kiloparsec scale jet maps suggest that the hot spot structures seen at
the periphery of FR II sources are associated with internal shocks and not
the bow shock.  This supports the use of parameters derived for hot spots, in
the construction of simple analytic models based on shock jump conditions,
and the Bernoulli equation.

Work is currently in progress to admit the consideration of time delays in
the radiation transfer calculations. Although we have argued here that for
the studies done to date these effects are unimportant, it will be crucial to
address the consequences of time delay for weak shocks that propagate at
approximately the speed of the underlying flow. A modification to the
hydrodynamic code is also being undertaken to include a passive magnetic
field.  This has major ramifications for the radiation transfer calculations,
as it will permit the production of maps in the Stokes parameters $Q$ and
$U$, for comparison with the lastest results from VLB polarimetry.

\acknowledgments

This work was supported by NSF grant AST 9120224 and by the Ohio Supercomputer
Center from a Cray Research Software Development Grant. 

\appendix
\section{The Relativistic Euler Equations}
 The hydrodynamic simulations are performed assuming axisymmetry, using as
physical variables the mass density $R$, the momentum density $M_{\rho}$ and
$M_{z}$, and the total energy density $E$ relative to the laboratory frame of
reference.  The gas is assumed to be inviscid and compressible with an ideal
equation of state with constant adiabatic index $\Gamma$. Using cylindrical
coordinates and defining the vector
\begin{equation}
 U =  (R,M_{\rho},M_{z},E)^{T} ,
\end{equation}
the two flux vectors
\begin{equation}
F^{\rho} = (R v^{\rho},M_{\rho} v^{\rho} + p, M_{z} v^{\rho},(E + 
p)v^{\rho})^{T } ,
\end{equation}
and
\begin{equation}
F^{z} = (Rv^{z},M_{\rho} v^{z},M_{z} v^{z} + p,(E + p)v^{z})^{T} ,
\end{equation}
and the source vector
\begin{equation}
S = (0,p/{\rho},0,0)^{T} ,
\end{equation}
the almost-conservative form of the equations is:
\begin{equation}
\frac{\partial{ U}}{\partial{t}} + \frac {1}{\rho}\frac{\partial}{\partial{\rho}
} (\rho F^{\rho}) + \frac{\partial}{\partial z} (F^{z}) = S .
\end{equation}
The pressure is given by the ideal gas equation of state
\begin{equation}
p = (\Gamma - 1) (e - n) ,
\end{equation}
where $e$ and $n$ are respectively the rest frame energy density and mass
density. In this work we use units in which the speed of light, $c$, is unity. 

The laboratory and rest frame variables are related via a Lorentz 
transformation:
\begin{equation}
R = \gamma n ,
\end{equation}
\begin{equation}
M_{\rho} = \gamma^2 ( e + p ) v^{\rho} , \ \ \ \
M_{z} = \gamma^2 ( e + p ) v^{z} ,
\end{equation}
\begin{equation}
E = \gamma^2 ( e + p ) - p ,
\end{equation}
where $\gamma = ( 1 - v^2 )^{-1/2}$ is the Lorentz factor and $v^2 = 
(v^{\rho})^2 + (v^{z})^2$.

In order to compute the pressure $p$ and sound speed $c_s$ we need the rest
frame mass density $n$ and energy density $e$.  However, these quantities are
nonlinearly coupled to the components of the velocity $v^{\rho}$ and $v^{z}$,
as well as to the laboratory frame variables $R$, $M_{\rho}$, $M_{z}$, and
$E$ via the Lorentz transformation given in equations (A7) to (A9).  When the
adiabatic index is constant it is possible to reduce the computation of $n$,
$e$, $v^{\rho}$, and $v^{z}$ to the solution of the following quartic
equation in the magnitude of the velocity $v$:
\begin{equation}
 \left[\Gamma v \left(E-Mv\right)-M\left(1-v^2\right)\right]^2 -\left(
 1-v^2\right)v^2\left(\Gamma-1\right)^2R^2=0 .
\end{equation}
Component velocities are then given by
\begin{equation}
v^{\rho} = {\rm sign}(M_{\rho}) v , \ \ \ \ 
v^{z} = M_{z} \frac{v^{\rho}}{M_{\rho}} .
\end{equation}
Then the quantities $e$ and $n$ can be found from the relations
\begin{equation}
e = E - M_{\rho} v^{\rho} - M_{z} v^{z} , \ \ \ \ n = \frac{R}{\gamma} .
\end{equation}

\section{An Analysis of the Transition between Emissivity and Doppler
Boost Domination}
 We wish to estimate the relative emissivity of the jet and of the post-bow
shocked ambient gas, to assess at what viewing angle the Doppler-boosted
former dominates the latter, thus causing the bow shock morphology to be lost
on maps with limited dynamic range. As noted in \S 3, the bow shock moves
forward at barely relativistic speed, thus we can use a nonrelativistic
description for its pressure distribution.

 In a first approximation the jet constitutes a blunt obstacle, and the bow
shock forms as a consequence of the flow over this structure, as the jet
propagates into the ambient medium. At a large axial distance ($r$) from the
obstacle, the `strength' of the bow shock -- defined in terms of the velocity
jump in the shock frame -- falls as $r^{-3/4}$ (Landau \& Lifshitz 1959).  We,
therefore, assume that
\begin{equation}
\frac{\Delta v}{\Delta v|_{\rm axis}}=\left(\frac{r_{\rm jet}}{r}\right)
^{3/4}, \ r>r_{\rm jet},
\end{equation}
the velocity (and pressure) jump being characterized by a single value in the
vicinity of the Mach disk.

 For a flow with adiabatic index $\Gamma$, the ratio of downstream to
upstream pressures in the shock frame is
\begin{equation}
\frac{p_d}{p_u}=\frac{2\Gamma{\cal M}_u^2-\left(\Gamma-1\right)}{\Gamma+1},
\end{equation}
while the corresponding velocity ratio is
\begin{equation}
\frac{v_d}{v_u}=\frac{2+\left(\Gamma-1\right){\cal M}_u^2}{\left(\Gamma+1
\right){\cal M}_u^2},
\end{equation}
where ${\cal M}_u$ is the upstream Mach number. Eliminating ${\cal M}_u$,
and expressing the velocity shift as $\Delta v=v_u-v_d$,
\begin{equation}
\Delta v=-v_u\left(\frac{4\Gamma/\left(\Gamma+1\right)}{\left(\Gamma+1
\right)\frac{p_d}{p_u}+\left(\Gamma-1\right)}-\frac{2}{\left(\Gamma+1
\right)}\right).
\end{equation}

On axis, the bow shock is strong, and $p_d>>p_u$, so that
\begin{equation}
\Delta v|_{\rm axis} \sim \frac{2v_u}{\left(\Gamma+1\right)} =
\frac{2v_{\rm bow}}{\left(\Gamma+1\right)},
\end{equation}
the final form arising from the fact that the bow propagates into a
stationary medium.

Note that in equation (B4),
$v_u$ is $\sin\chi\,v_{\rm bow}$, because only that
velocity component normal to the shock is modified by the shock. In
general the bow shock  forms an angle $\chi$ with respect to the axis of the
flow. Therefor, using
equations (B1) and (B5) for the variation of $\Delta v$ with $r$ and for
$\Delta v|_{\rm axis}$, we see that
\begin{equation}
\frac{p_d}{p_u}=\frac{2\Gamma/\left(1+\Gamma\right)}{\left[1-\frac{\left(
r_{\rm jet}/r\right)^{3/4}}{\sin\chi}\right]}-\frac{\left(\Gamma-1\right)}
{\left(\Gamma+1\right)}.
\end{equation}

Bow shocks appear to be approximately parabolic in form; fitting
a curve of form $r^n=-a\left(z-z_0\right)$ (where $z_0$ is the location of
the apex of the bow shock) to both the nonrelativistic and extreme
relativistic runs of Duncan \& Hughes (1994: runs A and D), demonstrate
quantitatively that $n\sim
2.2$. However, it is easily seen that using such a variation of $r(z)$ to
determine $\chi$ fails if $n>7/4$, for in that case, as $r\rightarrow\infty$,
$p_d/p_u\rightarrow-\infty$. The key to this apparent problem is that near to
the apex of the bow, $\chi\sim 90^{\circ}$, while for $r\gtrsim r_{\rm jet}$
the bow rapidly attains a constant angle ($\gtrsim 45^{\circ}$) with respect
to the axis; a parabola is a poor approximation globally. For $r\gtrsim
r_{\rm jet}$ it is appropriate to take $\sin\chi$ as a constant ${\cal
O}(1/\sqrt{2})$. We have checked the validity of this approximation by
computing $p_d/p_{\rm stag}$, with the stagnation pressure, ($p_{\rm stag}$),
computed from ${\cal M}_u$ following Landau \& Lifshitz; only for $r\lesssim
r_{\rm jet}$ does the ratio exceed unity, showing that the estimated pressure
downstream of the bow shock is unphysical.

To compare the post-bow shock emission and that of the jet, we note that as
the former is essentially nonrelativistic, and thus not Doppler boosted,
following the discussion of \S 3 the emissivity is $\varepsilon_{\rm bow}
\propto p_d^{\left(\alpha+3\right)/2}$, whereas the latter is
$\varepsilon_{\rm jet} \propto p_0^{\left(\alpha+3\right)} {\cal
D}^{\left(2+\alpha\right)}$, ${\cal D}$ the viewing angle-dependent Doppler
factor. As $p_0$ is the unshocked jet pressure, and the jet is taken to be
initially in pressure balance with the ambient gas, $p_u$ may be identified
with $p_0$. The common term in $p_0$ (i.e., $p_u$) means that we may compare
the jet and bow emissivities by comparing the values of $p_d/p_u$ and ${\cal
D}^{\left(2+\alpha\right)}$. This is done in Figure~18, which shows our
estimate of $p_d/p_u$ for the three relativistic runs of Duncan \& Hughes as
lines, and ${\cal D}^{\left(2+\alpha\right)}$ for these same runs as markers:
crosses for the $\gamma=5$ case, circles and squares for the $\gamma=10$
runs; $\chi$ was taken to be $55^{\circ}$. The jet radius is six units;
looking, for example, at three jet-radii, we see that the Doppler boosting of
the jet produces a comparable laboratory frame emissivity at an angle $\sim
18^{\circ}$ for the faster flows, and at $\sim 25^{\circ}$ for the slower
flow. Evidently, for viewing angles $\lesssim 30^{\circ}$, the Doppler boosted
jet will dominate emission from the bow more than a few jet-radii off axis.

\vfil
\eject

\figcaption[]{Schlieren-type images of laboratory frame density gradient
with: a. $\beta=0.3$ and $\Gamma =5/3$; b.  $\gamma=5$ and $\Gamma=5/3$; c.
$\gamma=10$ and $\Gamma=5/3$; d. $\gamma=10$  and $\Gamma=4/3$; e.  as d.
with inflow Lorentz factor modulated between 1 and 10 to induce
perturbations.
\label{fig1}}

\figcaption[]{Flux maps from the simulation shown in Figure~1a at viewing
angles:  a. $\theta=10^\circ$; ${\fam2  I}_{max}$=0.04773 b.
$\theta=30^\circ$; ${\fam2  I}_{max}$=0.03489 c.  $\theta=60^\circ$; ${\fam2
I}_{max}$=0.04478 and d. $\theta=90^\circ$; ${\fam2  I}_{max}$=0.05687.  The
contour levels are 5\%, 6.11\%, 7.45\%, 9.10\%, 11.12\%, 13.57\%, 16.57\%,
20.24\%, 24.71\%, 30.17\%, 36.84\%, 44.98\%, 54.93\%, 67.07\% and 81.90\% of
${\fam2  I}_{max}$
\label{fig2}}

\figcaption[]{Flux maps from the simulation shown in Figure~1b at viewing
angles:  a. $\theta=10^\circ$; ${\fam2  I}_{max}=1.833\times10^3$ b.
$\theta$=30$^\circ$; ${\fam2  I}_{max}=1.105\times10^3$   c.
$\theta$=60$^\circ$; ${\fam2  I}_{max}=7.181\times10^2$ and d.
$\theta$=90$^\circ$; ${\fam2  I}_{max}=6.770\times10^2$.  The contour levels
are the same as those used in Figure~2.
\label{fig3}}

\figcaption[]{Flux maps from the simulation shown in Figure~1c at viewing
angles:  a. $\theta$=10$^\circ$; ${\fam2  I}_{max}$=0.04773 b.
$\theta$=30$^\circ$; ${\fam2  I}_{max}$=0.03489 c.  $\theta$=60$^\circ$;
${\fam2  I}_{max}$=0.04478 and d. $\theta$=90$^\circ$; ${\fam2
I}_{max}=3.442\times10^4$.  The contour levels are the same as those used in
Figure~2.
\label{fig4}}

\figcaption[]{Flux maps from the simulation shown in Figure~1d at viewing
angles:  a. $\theta$=10$^\circ$; ${\fam2  I}_{max}=3.678\times10^6$ b.
$\theta$=30$^\circ$; ${\fam2  I}_{max}=1.331\times10^6$ c.
$\theta$=60$^\circ$; ${\fam2  I}_{max}=3.583\times10^5$ and d.
$\theta$=90$^\circ$; ${\fam2  I}_{max}=1.669\times10^5$. The contour levels
are the same as those used in Figure~2.
\label{fig5}}

\figcaption[]{Flux maps from the simulation shown in Figure~1e at viewing
angles:  a. $\theta$=10$^\circ$; ${\fam2  I}_{max}=8.050\times10^5$ b.
$\theta$=30$^\circ$; ${\fam2  I}_{max}=1.780\times10^5$ c.
$\theta$=60$^\circ$; ${\fam2  I}_{max}=6.273\times10^4$ and d.
$\theta$=90$^\circ$; ${\fam2  I}_{max}=3.796\times10^4$. The contour levels
are the same as those used in Figure~2.
\label{fig6}}

\figcaption[]{Linear gray scale plots, where darker color means higher
intensity, of a. the emissivity ($p^{2.75}$); ${\fam2
I}_{max}=1.110\times10^5$; \noindent the Doppler boosting (${\fam2
D}^{2.75}$) at b. $\theta$=30$^\circ$; ${\fam2  I}_{max}$=13.46 c.
$\theta$=60$^\circ$; ${\fam2  I}_{max}$=4.126 d. $\theta$=90$^\circ$;
${\fam2  I}_{max}$=2.447 of a slice through the perturbed jet.
\label{fig7}}

\figcaption[]{Plot of the flux (arbitrary units) verses angle for a region of
high Doppler boosting (solid line) and a region of the same size of high
emissivity offset a third of a jet radii from the flow (dashed line) for a
slice through the jet.
\label{fig8}}

\figcaption[]{Plot of flux (arbitrary units) verses angle summed for a line
of sight through a region of high Doppler boosting (solid line) and a line of
sight through a region of the same size of high emissivity offset from the
central axis (dashed line).
\label{fig9}}

\figcaption[]{Light curve of the time evolved perturbed jet.
\label{fig10}}

\figcaption[]{Light curve of BL Lac object 0735+178 from the UMRAO database.
\label{fig11}}

\figcaption[]{The periodogram analysis of central light curve shown in
Figure~10.
\label{fig12}}

\figcaption[]{Flux maps of the a. 11th b. 12th  c. 13th  and   d. 14th time 
slice of the evolving perturbed jet. ${\fam2  I}_{max}=1.560\times10^5$).
The contour levels are the same as those used in Figure~2.
\label{fig13}}

\figcaption[]{Component velocities.  The position of each component verses
time.
\label{fig14}}

\figcaption[]{Component flux evolution.  Flux verses how far the component
has traveled along the jet.
\label{fig15}}

\figcaption[]{A flux map of a two sided jet from the simulation shown in
Figure~1d at viewing angles a. $45^{\circ}$; ${\fam2
I}_{max}=5.383\times10^5$, b. $60^{\circ}$; ${\fam2
I}_{max}=3.288\times10^5$ and c. $75^{\circ}$; ${\fam2
I}_{max}=2.387\times10^5$.
\label{fig16}}

\figcaption[]{Schlieren-type image of the pressure gradient of the head of
the jet shown in Figure~1d. superimposed with white contours of the intensity
at $45^{\circ}$ of a slice through the same jet.  The arrow points to the
Mach disk.  The contour levels are 5.0\%, 6.7\%, 9.1\%, 12.3\%, 16.6\%,
22.4\%, 30.2\%, 40.7\%, 54.9\% and 75.1\% of the ${\fam2 I}_{max}=3.554
\times10^4$.
\label{fig17}}

\figcaption[]{Estimate of $p_d/p_u$ for the three relativistic runs of Duncan
\& Hughes as lines, and ${\cal D}^{\left(2+\alpha\right)}$ for these same
runs as markers:  crosses for the $\gamma=5$ case, circles and squares for
the $\gamma=10$ runs; $\chi$ was taken to be $55^{\circ}$.
\label{figB1}}

\end{document}